\documentclass[letter,scriptaddress,prl,showkeys,showpacs]{revtex4}
\usepackage{amsmath}
\usepackage{graphicx}
\usepackage{dcolumn}
\usepackage{bm}
\usepackage{graphicx}
\usepackage{epsfig}
\usepackage[usenames]{color}
\usepackage{tikz}
\usetikzlibrary{shapes}

\usepackage{setspace}
\usepackage[mathcal]{eucal}
\usepackage{subfig}
\usepackage{amsthm}
\usepackage{float}

\newcommand{\df}[2]{\displaystyle\frac{#1}{#2}}
\newcommand{\tf}[2]{\textstyle\frac{#1}{#2}}

\newcommand{\eps}{\varepsilon}

\newcommand{\be}{\begin{eqnarray}}
\newcommand{\en}{\end{eqnarray}}

\begin{document}
\title{Evolution of spherical cavitation bubbles:\\ 
parametric and closed-form solutions}
   \author{Stefan C. Mancas}
\email{mancass@erau.edu}
\affiliation{%
Hochschule M\"{u}nchen - Munich University of Applied Sciences, Germany
}%
\author{Haret C. Rosu}
\email{hcr@ipicyt.edu.mx}
\affiliation{IPICyT, Instituto Potosino de Investigacion Cientifica y Tecnologica,\\
Camino a la presa San Jos\'e 2055, Col. Lomas 4a Secci\'on, 78216 San Luis Potos\'{\i}, S.L.P., Mexico}


\begin{abstract}
We present an analysis of the Rayleigh-Plesset equation for a three dimensional vacuous bubble in water. In the simplest case when the effects of surface tension are neglected, the known parametric solutions for the radius and time evolution of the bubble in terms of a hypergeometric function are briefly reviewed. By including the surface tension, we show the connection between the Rayleigh-Plesset equation and Abel's equation, and obtain the parametric rational Weierstrass periodic solutions following the Abel route. In the same Abel approach, we also provide a discussion of the nonintegrable case of nonzero viscosity for which we perform a numerical integration.

\end{abstract}
\maketitle

\section{Introduction}

It is well established that the size evolution of unstable, spherical cavitation bubbles is governed by the Rayleigh-Plesset (RP) equation \cite{Lord,Prosp,Plass}
\be
\rho_w\Big(R\ddot{R}+\df{3}{2}\dot{R}^2\Big)=p-P_\infty-\frac 2 R\Big(\sigma+2 \mu_w\dot R\Big)~. \label{eq1}
\en
In (\ref{eq1}) $\rho_w$ is the density of the water, $R(t)$ is the radius of the bubble, $p$ and $P_\infty$ are respectively
the pressures inside the bubble and at large distance, $\sigma$ is the surface tension of the bubble, and $\mu_w$ is the
dynamic viscosity of water. In the simpler form with only the pressure difference in the right hand side, equation (\ref{eq1})
was first derived by Rayleigh in 1917 \cite{Lord} but it was only in 1949 that Plesset developed the full form of the equation
and applied it to the problem of traveling cavitation bubbles \cite{Plass}. In the second half of the last century a steady progress
has been achieved with driving forces from engineering, medical, sonoluminescence, microfluidics, and even pharmaceutical applications
of the cavitation phenomena \cite{Lugli}. In addition, the interest in the analytical and numerical solutions of cavitation dynamics remained considerable as these solutions can lead to a better control and understanding of the bubble collapse processes. In an effort
to discern the peculiar features of the usual three-dimensional collapse, Prosperetti \cite{Prosp} and more recently Klotz \cite{K} worked out generalizations to $N$-dimensional bubble dynamics. 

In this paper, we first review the analytic solutions of the RP equation
in terms of hypergeometric functions when the surface tension is neglected, and in terms of
Weierstrass elliptic functions when the surface tension is taken into account \cite{Kud,KS}. In
the latter case, we employ an Abel equation approach which is a novel mathematical way of looking
to the nonlinear evolution of cavitation bubbles.
On the other hand, when the viscous term is introduced, we show that an Abel equation with a non
constant invariant occurs. Since it is not yet known how to find analytical solutions to this equation
(if any), we resort to  numerical  integration of the stiff RP equation from $t=0$ to $t=t_c$, where $t_c$ is the time of collapse of the bubble.

\section{Size evolution and collapse of a spherical bubble without surface tension: Review of canonical results}

We first consider the idealized case whereby the viscosity of the water is neglected, since $\mu_w\ll 1$, and we further discard the effect of surface tension $\sigma$. For this case, there are also recent analytical approximations in \cite{Obr}, which are further discussed in \cite{Amore}, but here we are concerned with the standard results.

\begin{figure}[H]
\centering
\includegraphics[width=0.5\textwidth]{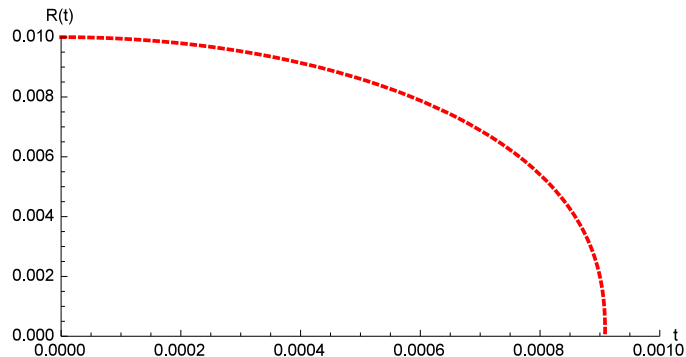}
\caption{\small{Radius of the bubble in the absence of surface tension according to equation \eqref{eq10a} for  $[\rho_w]=1000 ~ kg/m^3$,  $[R_0]=10^{-2}~ m$  and $[P_{\infty}]=101325 ~ Pa$.}}
\label{Fig1}
\end{figure}

Let us consider a vacuous $p=0$ bubble of radius $R$ which is surrounded by an infinite uniform incompressible fluid, such as water, that is at rest at infinity. We remark that `infinity' in the present context refers to distance far enough away from the initial position of the bubble, and we further  assume that the pressure at infinity is constant, $P_\infty=$ const.  Neglecting the body forces acting on the bubble, we have from equation \eqref{eq1}
\be
2 R\ddot{R}+3\dot{R}^2=-2\frac{P_\infty}{\rho_w}~.\label{eq2}
\en
Since $R^2\dot{R}$ is an integrating factor of \eqref{eq1} consequently we obtain by one quadrature
$$
R^3\dot{R}^2=-\df{2}{3}\df{P_\infty}{\rho_w}R^3+{\cal C}~.
$$
Using the initial conditions $R(0)=R_0$ and $\dot R(0)=0$ we find the integration constant to be
$$
{\cal C}=\df{2}{3}\df{P_\infty}{\rho_w}R_0^3~,
$$
and hence, we obtain
\be
\dot{R}^2=\df{2}{3}\df{P_\infty}{\rho_w}\left[\left(\df{R_0}{R}\right)^3-1\right]~.\label{eq3}
\en
Note that one can find a simple novel particular solution for $R(t)$ by  substituting \eqref{eq3} into \eqref{eq2} to obtain the Emden-Fowler equation
\be
\ddot R={\cal A} t^n R^m\label{eq5}
\en
with ${\cal A}=-\frac{3{\cal C}}{2}$, $n=0$, $m=-4$, and  particular solution
\begin{equation}
R_p(t)=\sqrt[5]{\frac{25{\cal C}}{4}}\left(t+\frac{2 R_0^{\frac 52}}{5 \sqrt{\mathcal C}}\right)^{\frac 2 5}=\sqrt[5]{\frac{25}{6}\frac{P_\infty R_0^3}{\rho_w}}\left(t+\frac{\sqrt 6}{5}\sqrt{\frac{\rho_w}{P_\infty}}R_0\right)^{\frac 2 5}~. \label{eq5a}
\end{equation}
However, this solution is obtained under the assumption of a nonzero integrating factor \cite{Obr} and therefore it fails to satisfy the second initial condition.
Other solutions that do not satisfy the initial conditions have been found previously by Amore and Fern\'andez \cite{Amore}.
As noticed in \cite{O-06}, equation~\eqref{eq3} can be also viewed as a conservation law for the dynamics of the radius of the bubble, since its kinetic energy can be expressed as
\be
2\pi \rho_w R^3\dot{R}^2=\tf{4}{3}\pi P_\infty(R_0^3-R^3)~.\label{eq5b}
\en
To proceed with the integration of equation \eqref{eq3} we will use the set of transformations as given by Kudryashov \cite{Kud}, namely
$R=S^\epsilon, dt=R^\delta d \tau$, where $\epsilon, \delta$ are constants that depend on the dimension of the bubble, and $S, \tau$ are the new dependent and independent variables, respectively.
Applying the transformations upon \eqref{eq3}, we obtain the new dynamics in $S$ and $\tau$
\be
S_\tau^2=\frac{2}{3}\frac{P_\infty}{\rho_w}\frac{1}{ \eps^2}(R_0^3~S^{-3 \eps}-1)S^{2+2\eps\delta-2\eps}~. \label{eq6}
\en
To find $S$, one sets $\eps=\frac{1}{N}$, and $\delta=N+1$, where $N=3$ is the dimension of the bubble, which will in turn reduce \eqref{eq6} to the simpler equation
\be
S_\tau=\sqrt {\frac{6 P_\infty}{\rho_w}}S\sqrt{R_0^3~S-S^2}~. \label{eq7}
\en
By integrating the above with $S(0)=R_0^3$ we obtain the rational solution
\be
S(\tau)=\frac{R_0^3}{{\cal B}\tau^2+1}~,\label{eq8}
\en
where for convenience we set ${\cal B}=\frac{9 {\cal C}}{4}R_0^3=\frac 3 2 \frac{P_\infty}{\rho_w}R_0^6.$ Once we determine $S$, we can  find the parametric solutions for the bubble radius $R(\tau)$ and evolution time of the bubble $t(\tau)$ \cite{Kud}
\begin{equation}
\begin{aligned}\label{eq9}
R(\tau)&= \frac{R_0}{({\cal B}\tau^2+1)^{\frac 1 3}}~,\\
t(\tau)&=R_0^4\int_0^{\tau} \frac{d \xi}{({\cal B} \xi^2+1)^{\frac 43}}~.
\end{aligned}
\end{equation}
The integral for the evolution of the time for bubble can be calculated analytically in terms of hypergeometric functions to give \cite{Kud}
\be
t(\tau)=\frac{R_0^4\tau}{2}\left[\frac{3}{\sqrt[3]{{\cal B} \tau^2+1}}-{}_2 F_1\left(\frac 1 2, \frac 1 3; \frac 3 2; -{\cal B} \tau^2\right)\right]=R_0^4 \tau{}_2 F_1\left(\frac 1 2, \frac 4 3; \frac 3 2; -{\cal B} \tau^2\right)~.\label{eq10}
\en
To achieve the time of collapse one needs to allow $\tau\rightarrow \infty$. This leads to $\lim_{\tau \rightarrow \infty} t(\tau)=0.000908681$ if we use, in S.I. units, $[\rho_w]=1000 ~ kg/m^3$,  $[R_0]=10^{-2}~ m$  and $[P_{\infty}]=101325 ~ Pa$.
We point out that the same asymptotic value for the collapse time has been also obtained, albeit using a different approach,  by Obreschkow {\em et al}. \cite {Obr} and it is also obtained from Eq.~\ref{eq12} below by direct integration.
Once we solve for $\tau$ as a function of $R$ from the first equation of \eqref{eq9}, and substituting it into the second equation of \eqref{eq9} we obtain the closed-form solution
\begin{widetext}
\be
t(R)=R_0\sqrt{\frac 2 3\frac{\rho_w}{ P_\infty}}\sqrt{\Big(\frac{R_0}{R}\Big)^{3}-1}~{}_2 F_1\left(\frac 1 2,\frac 4 3; \frac 3 2; 1-\Big(\frac{R_0}{R}\Big)^3\right)~
\label{eq10a}
\en
which is plotted as $R(t)$ in Fig.~\ref{Fig1}.
\end{widetext}

Next, we will find the time for the total collapse $t_c$ of the bubble by  integration of  equation (\ref{eq3}), which yields
\be
t_c=\frac{1}{R_0^{3/2}}\sqrt{\df{3}{2}\df{\rho_w}{P_\infty}}\int_0^{R_0}\df{R^{3/2}}{\sqrt{1-\Big(\frac{R}{R_0}\Big)^3}}\,dR~.\label{eq11}
\en
This is just the $N=3$ case of the general $N$-dimensional formula given in \cite{K}.
If we let $R=R_0\sin^{2/3}\theta$, where $\theta\in[0, \pi/2]$ then the integral (\ref{eq11}) transforms to
$$
t_c=\df{2R_0}{3}\sqrt{\df{3}{2}\df{\rho_w}{P_\infty}}\int_0^{\pi/2}\sin^{2/3}\theta\,d\theta,
$$
which by comparing with the integral relation for Beta function, namely,
$$
{\rm B}(m,n)=2\int_0^{\pi/2}\cos^{2m-1}\theta\sin^{2n-1}\theta\,d\theta
$$
leads to
\be \label{eq12}
t_c=\df{R_0}{3}\sqrt{\df{3}{2}\df{\rho_w}{P_\infty}}{\rm B}\left(\frac{1}{2},\frac{5}{6}\right)=
R_0\df{\Gamma(\frac{5}{6})}{\Gamma(\frac{4}{3})}\sqrt{\df{\pi\rho_w}{6P_\infty}}=0.914681~R_0\sqrt{\frac{\rho_w}{P_\infty}}~.
\en
This solution can be also obtained from \eqref{eq5} by using the parametric solution $2.3.1-2.2$ from Polyanin\rq{}s book \cite{Pol}.
If we insert in \eqref{eq12} the same S.I. numbers as used in the hypergeometric asymptotics, we obtain again $t_c=0.000908681~ sec$. 

\section{Surface tension included via Abel\rq{}s equation}
When the surface tension term added, equation \eqref{eq1} can be written in the form
\be
\ddot R+\frac{3}{2R}\dot R^2+\frac{K_1}{R}+\frac{K_2}{R^2}=0~,\label{eq12a}
\en
where we define $K_1=\frac{P_\infty-P}{\rho_w}$ and $K_2=\frac{2\sigma}{\rho_w}$.

Proceeding as in \cite{Man3}, we first show that solutions to a general second order ODE of type
\begin{equation}\label{eq14}
\ddot R+f_2(R)\dot R  +f_3(R)+f_1(R) \dot R^2+f_0(R) \dot R^3=0
\end{equation}
 may be obtained via the solutions to Abel's equation (\ref{eq15}) of the first kind (and vice-versa)
\begin{equation}\label{eq15}
\frac{dy}{dR}=f_0(R)+f_1(R)y+f_2(R)y^2+f_3(R)y^3
\end{equation}
using the substitution
\begin{equation}\label{eq16}
\dot R=\eta(R(t))~,
\end{equation}
which turns (\ref{eq14}) into the Abel equation of the second kind in canonical form
\begin{equation}\label{eq17}
\eta \dot \eta+f_3(R)+f_2(R)\eta+f_1(R)\eta^2+f_0(R)\eta^3=0~.
\end{equation}
Moreover, via the inverse transformation
\begin{equation}\label{eq19}
\eta(R(t))=\frac{1}{y(R(t))}
\end{equation}
of the dependent variable, equation (\ref{eq17}) becomes (\ref{eq15}).
The invariant of Abel\rq{}s equation, see \cite{kam}, can be written as
\be
\Phi(R)=\frac 1 3 \left(\frac{df_2}{dR}f_3-f_2\frac{df_3}{dR}-f_1f_2f_3+\frac 2 9 f_2^3\right)\label{eq19a}
\en
and when is a constant is an indication that Abel\rq{}s equation is integrable.

By identification of equation \eqref{eq12a} with \eqref{eq14}, we see that $f_1(R)=\frac{3}{2R}$, $f_2(R)=f_0(R)=0$, and $f_3(R)=\frac{K_1}{R}+\frac{K_2}{R^2}$, and hence the Kamke invariant is $\Phi(R)=0$, therefore
Abel\rq{}s equation \eqref{eq15} becomes the Bernoulli equation
\be
\frac{dy}{dR}=f_1(R)y+f_3(R)y^3\label{eq20}
\en
 which, by one quadrature, has the solution
 \be
y(R)=\frac{\pm \sqrt 3 R^{\frac 32}}{\sqrt{3{\cal D}-2 K_1R^3-3 K_2 R^2}}~.\label{eq21}
\en
By using equations \eqref{eq19} and \eqref{eq16}, we obtain
\be
\dot R^2=\frac{3 {\cal D}-2 K_1R^3-3 K_2 R^2}{3 R^3} \label{eq22}
\en
and via the same initial conditions we obtain the integration constant
$$
{\cal D}=\frac{R_0^2}{3}(3K_2+2K_1R_0)~,
$$
which gives
\be
\dot R^2=\frac{2K_1(R_0^3-R^3)+3K_2(R_0^2-R^2)}{3 R^3}~. \label{eq23}
\en
Notice that when $\sigma=0 \rightarrow K_2=0$, and $p=0\rightarrow K_1=\frac{P_\infty}{\rho_w}$, then the above becomes equation \eqref{eq3}. The new energy with surface tension is
\be
2\pi \rho_w R^3\dot{R}^2=\frac{4\pi}{3}\left[P_\infty(R_0^3-R^3)+3\sigma(R_0^2-R^2)\right]~.\label{eq24}
\en
Thus, at the time of collapse we have $\frac{4 \pi}{3}R_0^2(P_\infty R_0+3 \sigma)= 0.42443J$ for a surface tension $[\sigma]=10^{-3}~N/m$.

\begin{figure}[H]
\centering
\includegraphics[width=0.35\textwidth]{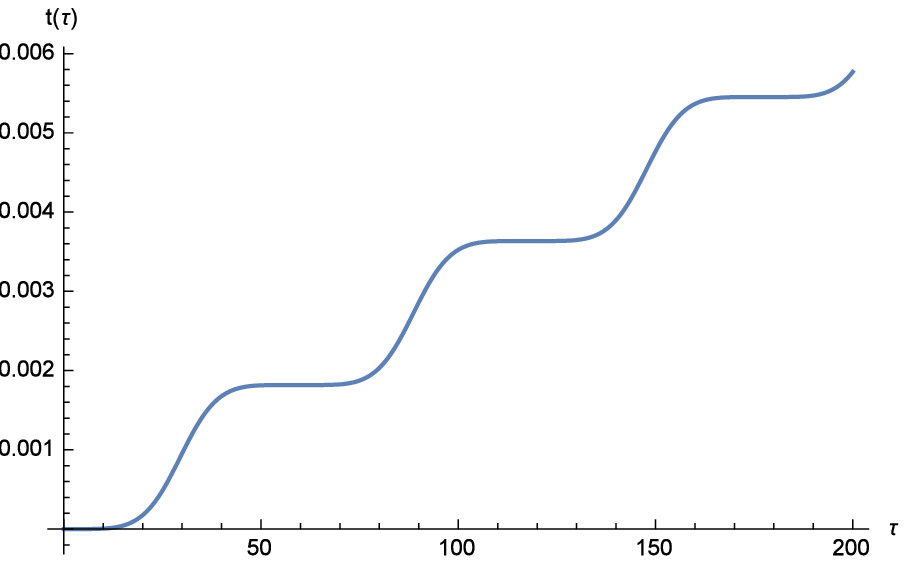}\\
\includegraphics[width=0.35\textwidth]{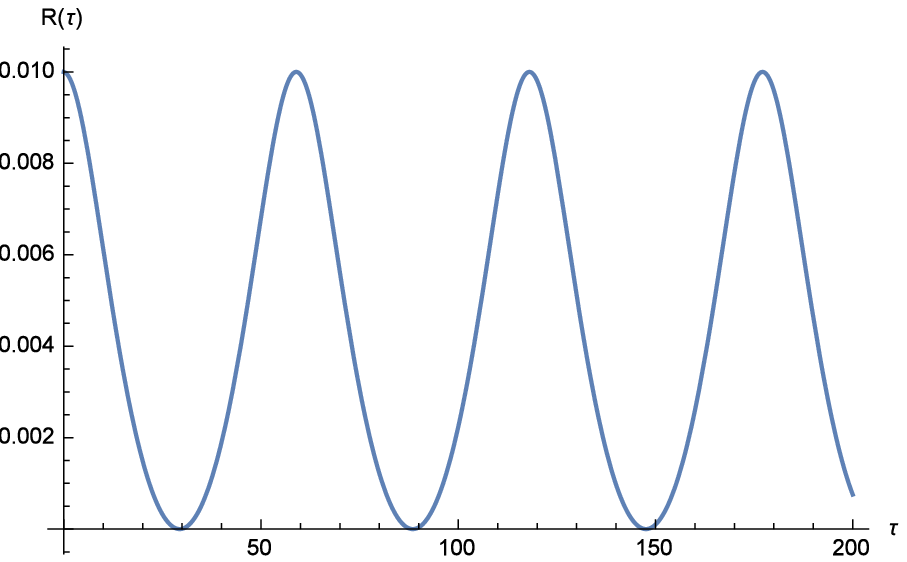}\\
\includegraphics[width=0.35\textwidth]{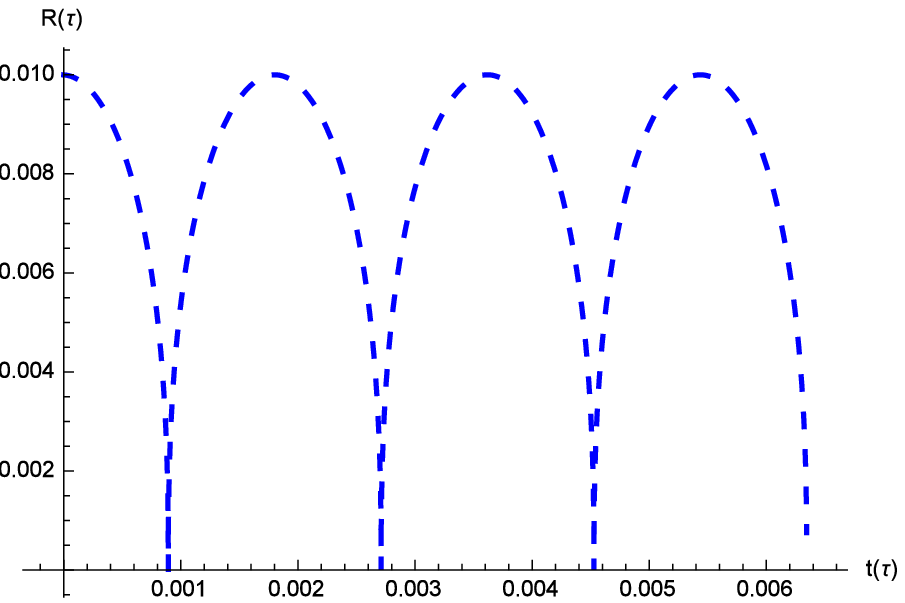}
\caption{\small{Parametric solutions for time of evolution (top), radius of the bubble (middle), and radius vs time (bottom) from equation \eqref{eq31} when surface tension is present for $[\rho_w]=1000 ~ kg/m^3$,  $[R_0]=10^{-2}~ m$,  $[P_{\infty}]=101325 ~ Pa$ and $ [\sigma]=10^{-3}~ N/m.$}}
\label{Fig2}
\end{figure}

To integrate equation \eqref{eq23},  first let us write it in a more convenient way, as
\be
\dot R^2=\frac{a_3}{R^3}+\frac{a_1}{R}+a_0\label{eq24a}
\en
with coefficients defined as  $a_3=R_0^2\left(K_2+\frac{2K_1}{3}R_0\right)$, $a_1=-K_2$, and $a_0=-\frac{2K_1}{3}$, and we will use the same set of  transformations, namely
$R=S^\epsilon$, $dt=R^\delta d \tau$ which give in turn
\be
S_\tau^2=\frac{S^{2+2 \eps \delta}}{\eps^2}(a_0S^{- 2\eps}+a_1S^{- 3\eps}+a_3S^{- 5\eps})~,  \label{eq24b}
\en
where now we set $\epsilon=-3/N=-1$, and $\delta=\frac{N+1}{2}=2$.
Thus, we obtain the Weierstrass elliptic equation
\be
S_\tau^2=a_0+a_1S+a_3S^3 \label{eq25}
\en
which in standard form is
\begin{equation}\label{eq26}
\wp_{\tau}\,^2=4 \wp^3-g_2 \wp -g _3
\end{equation}
via the linear substitution \cite{Whi}
\be \label{eq27}
S(\tau)=\frac{4}{a_3}\wp(\tau+29.709;g_2,g_3)~.
\en
The  germs of the Weierstrass function $g_2,g_3$ are given by
\begin{equation}\label{eq29}
\begin{aligned}
g_2&=-\frac{a_1a_3}{4}=\frac{K_2R_0^2}{4}\big(K_2+\frac {2K_1R_0}{3}\big)\\
g_3&=-\frac{a_0 a_3^2}{16}=\frac{K_1R_0^4}{24}\big(K_2+\frac {2K_1R_0}{3}\big)^2~.
\end{aligned}
\end{equation}
Substituting $K_1$ and $K_2$ into the germs, the solution to equation \eqref{eq25}  becomes
\begin{equation}\label{eq30}
S(\tau)=\frac{6 \rho_w}{R_0^2(P_\infty R_0+3 \sigma)}\wp(\tau+29.709;g_2,g_3)~,
\end{equation}
where $g_2=3.37751 \cdot 10^{-11}~m^8/sec^4$, $g_3=1.92645 \cdot 10^{-8}~ m^{12}/sec^6$, and the constant in the front of Weierstrass function from \eqref{eq30} takes the value of $59215.2 ~ \sec^2/m^5$.
Once $S$ is known we can find the parametric solutions for the radius and time of the bubble with surface tension as
\begin{widetext}
\be
\begin{aligned}\label{eq31}
R(\tau)&= \frac{1}{S(\tau)}=\frac{R_0^2(P_\infty R_0+3 \sigma)}{6 \rho_w}\frac {1}{\displaystyle{\wp} \left(\tau+29.709;\frac{R_0^2 \sigma}{3 \rho_w^2}(P_\infty R_0+3 \sigma),\frac{R_0^4 P_\infty}{54 \rho_w^3}(P_\infty R_0+3 \sigma)^2\right)}\\
t(\tau)&=\int_0^\tau \frac{d \xi}{S(\xi)^2}=\frac{R_0^4(P_\infty R_0+3 \sigma)^2}{36 \rho_w^2}\displaystyle \int_0^\tau \frac{d \xi}{\displaystyle {\wp}\left( \xi;\frac{R_0^2 \sigma}{3 \rho_w^2}(P_\infty R_0+3 \sigma),\frac{R_0^4 P_\infty}{54 \rho_w^3}(P_\infty R_0+3 \sigma)^2\right)^2}~.
\end{aligned}
\en
Related plots are presented in Fig. \ref{Fig2}.
\end{widetext}

 \begin{figure}[H]
\centering
\includegraphics[width=0.4\textwidth]{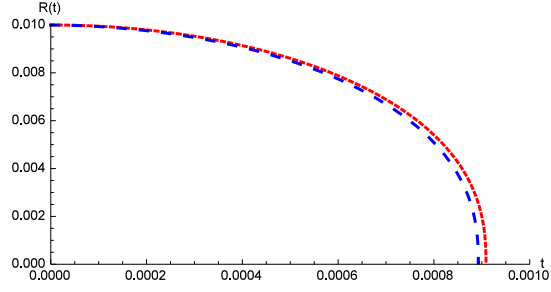}
\caption{\small{Comparison between the exact solutions without (dotted curve) and with  surface tension (dashed curve). Notice that when we have surface tension the time of collapse is smaller. }}
\label{Fig3}
\end{figure}
\section{The Rayleigh-Plesset equation with viscosity}

When we include the viscosity, equation \eqref{eq1} reads
\be
\ddot R+f_2(R) \dot R+f_1(R)\dot R^2+f_3(R)=0~,\label{eq35}
\en
where $f_2(R)=\frac{K_3}{R^2}$, and $K_3=\frac{4 \mu_w}{\rho_w}$ is a constant, and $f_1(R), f_3(R)$ being the same functions as before. Thus, Abel\rq{}s equation \eqref{eq15} becomes
\begin{equation}\label{eq36}
\frac{dy}{dR}=f_1(R)y+f_2(R)y^2+f_3(R)y^3~.
\end{equation}
Now, we also find the Kamke invariant according to equation \eqref{eq19a} and we obtain
\be
\Phi(R)=\frac{K_3[4K_3^2-9R(3K_2+5K_1R)]}{R^6}~.
\label{eq36a}
\en
In terms of the physical variables of the system, the invariant is
\be
\Phi(R)=\frac{2 \mu_w[64 \mu_w^2-9\rho_wR(6 \sigma+5 P_\infty R)]}{27 \rho_w^3 R^6}\label{eq36b}
\en
and because is not a constant, we will try to reduce \eqref{eq36} using the Appell invariant instead.

First, we will eliminate the linear term via the transformation $y(R)=R^{\frac 3 2}z(R)$ to obtain the reduced Abel equation
\begin{equation}\label{eq37}
\frac{dz}{dR}=h_2(R)z^2+h_3(R)z^3~,
\end{equation}
where $h_2(R)=\frac{K_3}{\sqrt R}$, and $h_3(R)=(K_1R+K_2)R$.

\begin{figure}[H]
\centering
\includegraphics[width=0.4\textwidth]{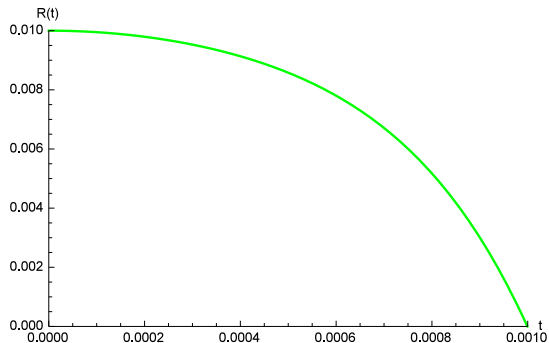}
\caption{\small{Numerical solution for the RP equation \eqref{eq35} when surface tension and viscosity are both present for  $[\rho_w]=1000 ~ kg/m^3$,  $[R_0]=10^{-2}~ m$,  $[P_{\infty}]=101325 ~ Pa$, $ [\sigma]=10^{-3}~ N/m $ and $[\mu_w]=1.002 \cdot 10^{-3} kg/(m \cdot sec)$.}}
\label{Fig4}
\end{figure}

According to the book of Kamke \cite{kam}, for equations of the type \eqref{eq37} for which there is no constant invariant one should change the variables according to
\begin{equation}\label{eq38}
\begin{aligned}
	z(R)&=\hat{z}(\zeta(R)) ~,\\
	 \zeta(R) &= \int {h_2(R)dR}=2K_3\sqrt R~,\\
\end{aligned}
\end{equation}
which lead to the canonical form
\begin{equation}\label{eq40}
\frac{d {\hat{z}}}{d \zeta}=\hat{z}^2+\Psi(\zeta)\hat{z}^3~,
\end{equation}
where
\begin{equation}\label{eq41}
\Psi(\zeta)=\frac{h_3\big(R(\zeta)\big)}{h_2\big(R(\zeta)\big)}=\zeta^3(b_3+b_5\zeta^2)
\end{equation}
is the Appell invariant, and the constants $b_i$ are
 \begin{equation}\label{eq42}
\begin{aligned}
	b_3&=\frac{K_2}{8 K_3^4}=\frac{\sigma \rho_w^3}{2^{10} \mu_w^4}~,\\
	 b_5&=\frac{K_1}{32 K_3^6}=\frac{P_\infty \rho_w^5}{2^{17} \mu_w^6}~.\\
\end{aligned}
\end{equation}
Choosing a dynamic viscosity of water $[\mu_w]=1.002 ~cP=1.002 \cdot 10^{-3} ~kg/(m \cdot sec)$ the $b_i$ constants take the values of $b_3=9.68789 \cdot 10^{14}~sec^2/m^5$, while $b_5=7.63836\cdot 10^{32}~sec^4/m^{10}$. The units of $\zeta$ are $m^{\frac 5 2}/sec$.
This Abel equation is not integrable through quadratures, but numerically we integrate the RP  equation \eqref{eq35}, from $t=0$ to  the point of stiffness which is the point in time where the bubble collapses, see the numerical  solution on  Fig. \ref{Fig4}.

 \begin{figure}[H]
\centering
\includegraphics[width=0.4\textwidth]{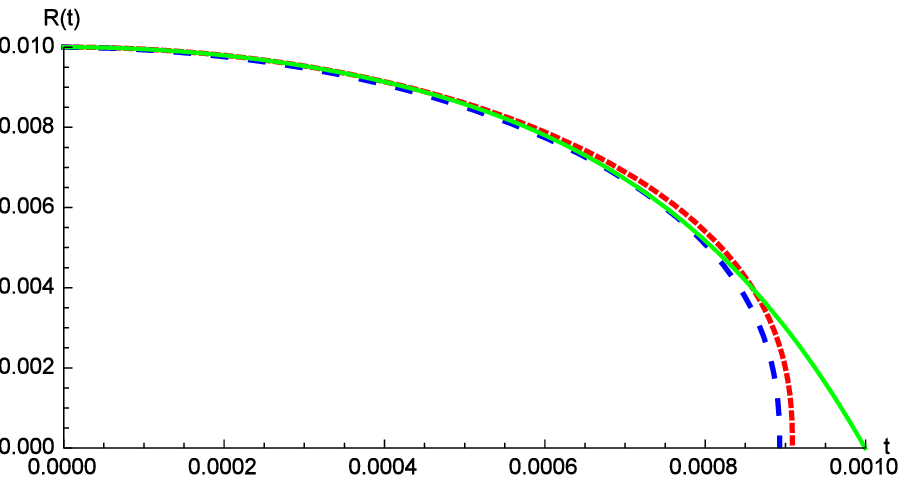}
\includegraphics[width=0.45\textwidth]{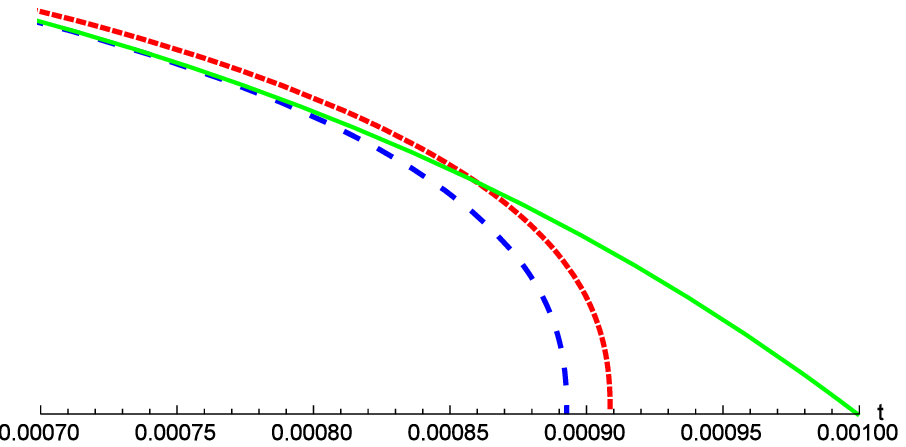}
\caption{\small{Comparison between the numerical solution (continuous curve) of  Fig.  \ref{Fig4} and the exact solutions  of Fig.  \ref{Fig3} .  As one can see, the bubbles collapse fastest when there is surface tension (dashed curve) followed  by the no surface tension and no viscosity case (dotted curve). The slowest time is achieved when we have viscosity together with surface tension (continuous curve).  Right panel is the zoomed in version of the figure in the left panel.}}
\label{Fig5}
\end{figure}

\section{Conclusion}
In this work, we have considered the RP equation for the size evolution of a bubble in water.
In the first part of the paper, we have surveyed the standard results in the absence of surface tension but in a different way from those usually pursued in the literature. We have obtained the closed form hypergeometric solutions of Kudryashov and Sinelshchikov although in a different but equivalent form. From an Emden-Fowler form of the RP equation in this case we have also obtained a particular solution which however does not satisfy the second initial condition. In the presence of surface tension and viscosity, we have employed a new approach based on Abel's equation. When only the surface tension is included, we have obtained the known parametric rational Weierstrass solutions, whereas when viscosity is added, the corresponding Abel equation does not have a constant invariant, which explains the nonintegrability in this case.  A numerical integration obtained from this nonintegrable Abel route is presented graphically.

\newpage

\noindent {\bf Acknowledgment}

\noindent This research was partially supported by internal funding from Embry-Riddle Aeronautical University. We thank the reviewers for their appropriate remarks which improved the quality of  the paper.

\end{document}